\input epsf.tex
\documentstyle[aps,multicol,prb]{revtex}
\begin{document}
\draft
\title{Anisotropic Superexchange for nearest and next nearest coppers\\
 in chain,
ladder and lamellar cuprates}
\author{Sabine Tornow, O. Entin-Wohlman and Amnon Aharony}
\address{School of Physics and Astronomy, Raymond and Beverly Sackler Faculty 
of Exact Sciences,\\
Tel Aviv University, Tel Aviv 69978, Israel}
\date{\today}
\maketitle
\begin{abstract}
We present
a detailed calculation of the magnetic couplings
between nearest--neighbor and next--nearest--neighbor
coppers in the edge--sharing geometry, ubiquitous in many
cuprates. In this geometry, the interaction between
nearest neighbor coppers is mediated via two oxygens, and the Cu--O--Cu
angle is close to 90$^\circ$.
The derivation is based on a perturbation expansion
of a general Hubbard Hamiltonian, and produces numerical estimates for the
various magnetic energies. In particular we find the dependence of the
anisotropy energies on the angular deviation away from the 90$^{\circ}$
geometry of the Cu--O--Cu bonds. 
Our results are required for the correct analysis of the
magnetic structure of various chain, ladder and lamellar cuprates.
\end{abstract}
\begin{multicols}{2}

\section{Introduction}

The magnetic interactions in the copper oxides
are believed to be governed by kinetic superexchange through the intervening
oxygens. In tetragonal symmetry, one may view the CuO planes as
consisting of clusters of four oxygens, forming a square whose
center is occupied by a copper ion. 
These squares can be lined up along their edges, and then the
nearest neighbor (NN) Cu--O--Cu bond makes an almost 90$^{\circ}$
angle. Another ubiquitous configuration is formed when the squares are
connected along their corners, in which case the nearest neighbor
Cu--O--Cu bond is linear, having an angle of 180$^{\circ}$. Typical
examples for edge--sharing compounds are La$_{6}$Ca$_{8}$Cu$_{24}$O$_{41}$
\cite{carter}
(where the angle is 91$^{\circ}$) and CuGeO$_{3}$ (where the
angle is $\approx 98^{\circ}$). \cite{hase} Corner--sharing 
configurations characterize
the copper oxide planes in the parent compounds of the high--$T_{c}$
cuprates, and the chains in Sr$_{2}$CuO$_{3}$ \cite{kojima}
and SrCuO$_{2}$. \cite{matsuda2}
Some compounds include both types of bonds, for example 
Sr$_{2}$Cu$_{3}$O$_{4}$Cl$_{2}$. \cite{yamada} The various Cu--Cu bond geometries
in this material are the same as those for the nearest--neighbors and
next--nearest--neighbors (NNN) in the chains in
Sr$_{14}$Cu$_{24}$O$_{41}$ \cite{matsuda}
and the interladder bonds in
Sr$_{n-1}$Cu$_{n+1}$O$_{2n}$. \cite{geillo}

The magnitude and the sign of the magnetic interactions
in the two types of bonds are expected to be quite different.
According to the so--called Goodenough--Kanamori--Anderson (GKA) rules,
\cite{GKA} the {\it leading} isotropic superexchange of a
180$^{\circ}$
bond between two magnetic ions
with partially filled d shells is strongly antiferromagnetic,
while the leading order of a 90$^{\circ}$ 
superexchange is ferromagnetic, 
and much weaker.
In the Cu--O case, the reason for this is that 
for the corner--sharing geometry, 
the 2p$_{\sigma}$ orbital hybridizes with the
two neighboring Cu ions, yielding a significant contribution to the kinetic
superexchange (which is antiferromagnetic). In contrast,
in the edge--sharing
configuration the 2p$_{\sigma}$ orbital on the oxygen, which
hybridizes with a 3d orbital on one copper, is almost orthogonal
to that 3d orbital on the nearest--neighbor Cu ion, thus blocking the
antiferromagnetic superexchange via a single oxygen. The leading magnetic
coupling in this case is given by the next order perturbation terms,
and is therefore weaker and of the opposite sign. 

Higher--order perturbation terms also determine
the magnetic {\it anisotropies}, in both types of bonds. These
anisotropies are responsible for various observable quantities, like
the gaps in the spin wave spectrum, the spin orientations in space, etc,
and hence are of much interest.
The magnetic couplings of the linear Cu--O--Cu bond were investigated in great
detail (see Refs. \onlinecite{yildirim} and \onlinecite{stein} and references
therein), yielding the in--plane and the out--of--plane gaps
of the family of compounds with structures similar to that
of La$_{2}$CuO$_{4}$, as well as the antisymmetric Dzyaloshinskii--Moriya
interaction in the orthorhombic phase. In particular, the
role of the on--site Coulomb exchange and the spin--orbit interaction in
producing the various anisotropies was clarified in tetragonal
and orthorhombic symmetries. 

\begin{figure}
\epsfxsize=6cm
\centerline{{\epsffile{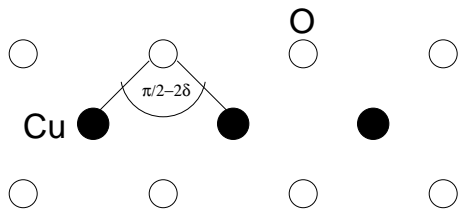}}}
\vspace{0.1cm}
\caption{Edge--sharing Cu--O configuration. The angle of the  Cu--O--Cu
bond is $\pi/2-2\delta$. Open circles denote oxygens and
black circles are the Cu's.}
\label{fig1}
\end{figure}

In the almost--90$^{\circ}$ bond (see Fig. \ref{fig1}) 
the leading order magnetic exchange is small. Therefore, 
higher--order perturbation processes, as well as details of the structure like
the presence of side groups \cite{geertsma,braden} have a significant
contribution. For the same reason,
next--nearest--neighbor Cu--Cu couplings are expected to be much
more important than in the case of the linear Cu--Cu bond configuration.

Previous discussions of this geometry include an analysis of the
dependence of the leading nearest--neighbor {\it isotropic} coupling
 on the small angular deviation away from 
90$^{\circ}$, $\delta $, \cite{geertsma,braden,gopalan,mizuno,muller}
which has been found to be
dominated by the on--site Coulomb exchange interaction
on the oxygens,\cite{gopalan} and the non--local exchange between 
the coppers and the oxygens.\cite{geertsma,braden} The magnetic
{\it anisotropies} have been calculated only for the
nearest--neighbor, strictly 90$^{\circ}$--bond, by Yushankhai and Hayn.
\cite{yushankhai} We compare below their results with ours.

The aim of this paper is to present a detailed calculation of
both the isotropic and the anisotropic magnetic interactions 
in the nearly--90$^{\circ}$ configuration shown in Fig. \ref{fig1}, for
nearest--neighbor and next--nearest--neighbor copper ions.
Our calculation is based on the perturbation expansion of a Hubbard
model around the half-filled ground state, in which there is a single 3d hole
on each copper ion, and the oxygen 2p states are completely full. 
Since the spin of the Cu hole is arbitrary, this ground state is 
2$^{\rm N}$--fold degenerate, where N is the number of copper ions. The
superexchange magnetic Hamiltonian is obtained as the effective interaction
within this degenerate manifold. The microscopic Hamiltonian we consider
includes hopping between all orbitals
on the copper d--states and on the oxygen p--states, 
and between the p states
on neighboring oxygens. The Hamiltonian also contains the spin--orbit
interactions on the copper. Those on the oxygen are much weaker and are 
therefore neglected. We also include all local Coulomb interactions on the
copper and on the oxygen, and the non--local Coulomb exchange between the
copper and the oxygen. 

General expressions for the effective magnetic
Hamiltonian, based on this micorscopic Hamiltonian have been derived before.
\cite{yildirim,stein} In order to keep the paper self--contained
we reproduce in Sec. II the main steps of the derivation. 
We find in Sec. II that the magnetic Hamiltonian has the form
\begin{equation}
{\cal H}=\sum_{\langle ij\rangle}\sum_{\mu}\bigl (
J_{\rm NN}^{\mu}S_{i}^{\mu}S_{j}^{\mu}+J_{\rm NNN}^{\mu}S_{i}
^{\mu}S_{j}^{\mu}\bigr ), \label{HM}
\end{equation}
in which $\mu $ denotes the Cartesian component of the spin and $J_{\rm NN}$ and
$J_{\rm NNN}$ are the magnetic couplings between nearest--neighbors and
next--nearest--neighbors, respectively. It is convenient to 
define the coordinate
system for the spin components such that the $x$ and $y$ directions
are in the Cu--O plane, along the bonds between the coppers, and the
$z$ direction is perpendicular to the plane. 
The leading magnetic coupling for both NN and NNN is
\begin{equation}
J^{\rm av}=\frac{J^{x}+J^{y}+J^{z}}{3}.
\end{equation}
The anisotropic couplings are then naturally given by $J^{\rm op}$,
for the out--of--plane anisotropy, and $J^{\rm pd}$, for the
in--plane one:
\begin{equation}
J^{\rm op}=J^{z}-\frac{J^{x}+J^{y}}{2},\ \ J^{\rm pd}=\frac{J^{x}-J^{y}}{2}.
\end{equation}
(The notation ``pd" stands for pseudo--dipolar, see 
Ref.\onlinecite{geillo}.)

\begin{figure}
\epsfxsize=8cm
\centerline{{\epsffile{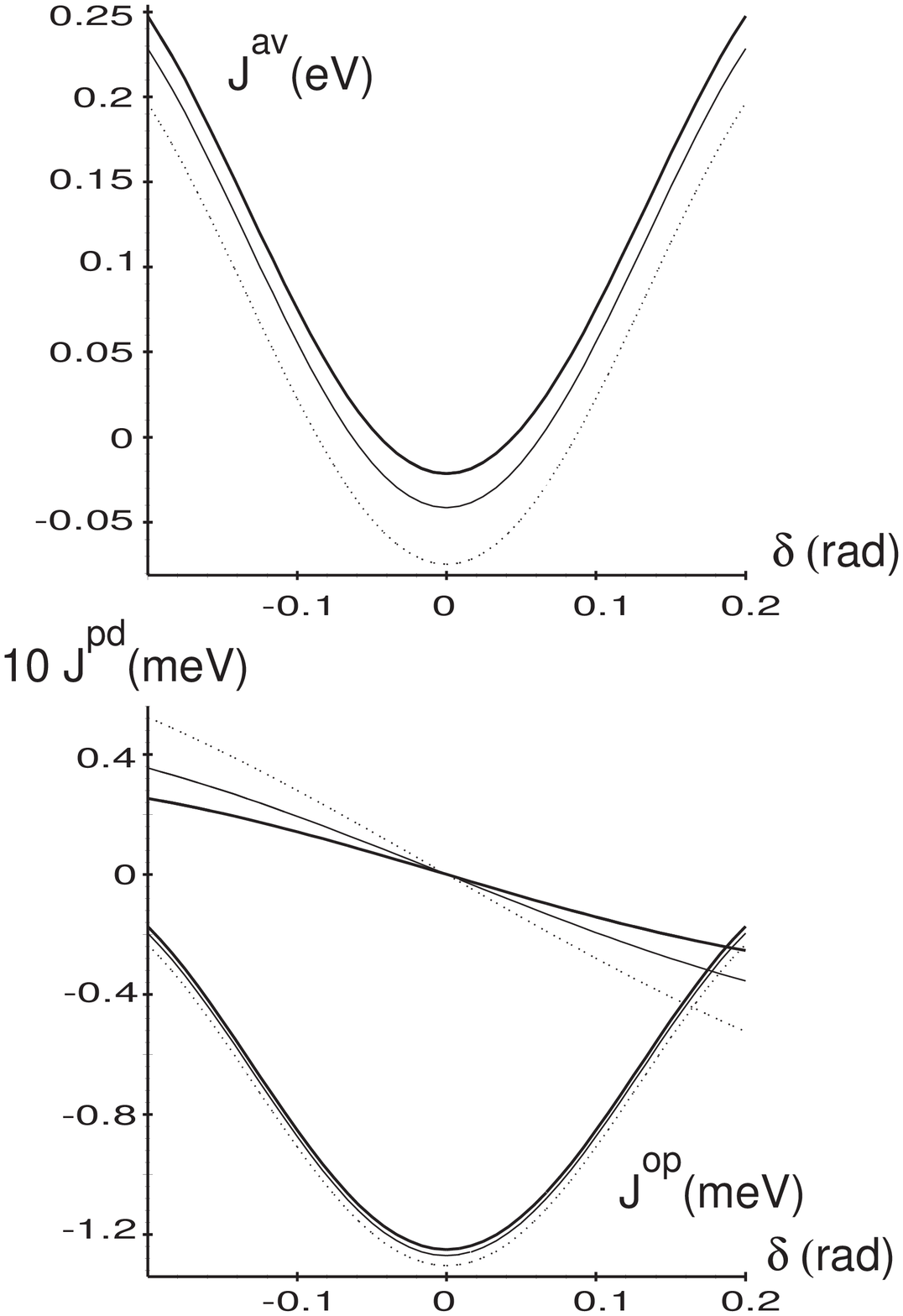}}}
\vspace{2cm}
\caption{ The nearest--neighbor magnetic couplings $J_{\rm NN}$
for K=0.02 eV (bold solid line), K=0.05 (solid line) and
K=0.1 eV (dashed line).}
\label{fig2}
\end{figure}

\begin{figure}
\epsfxsize=8cm
\centerline{{\epsffile{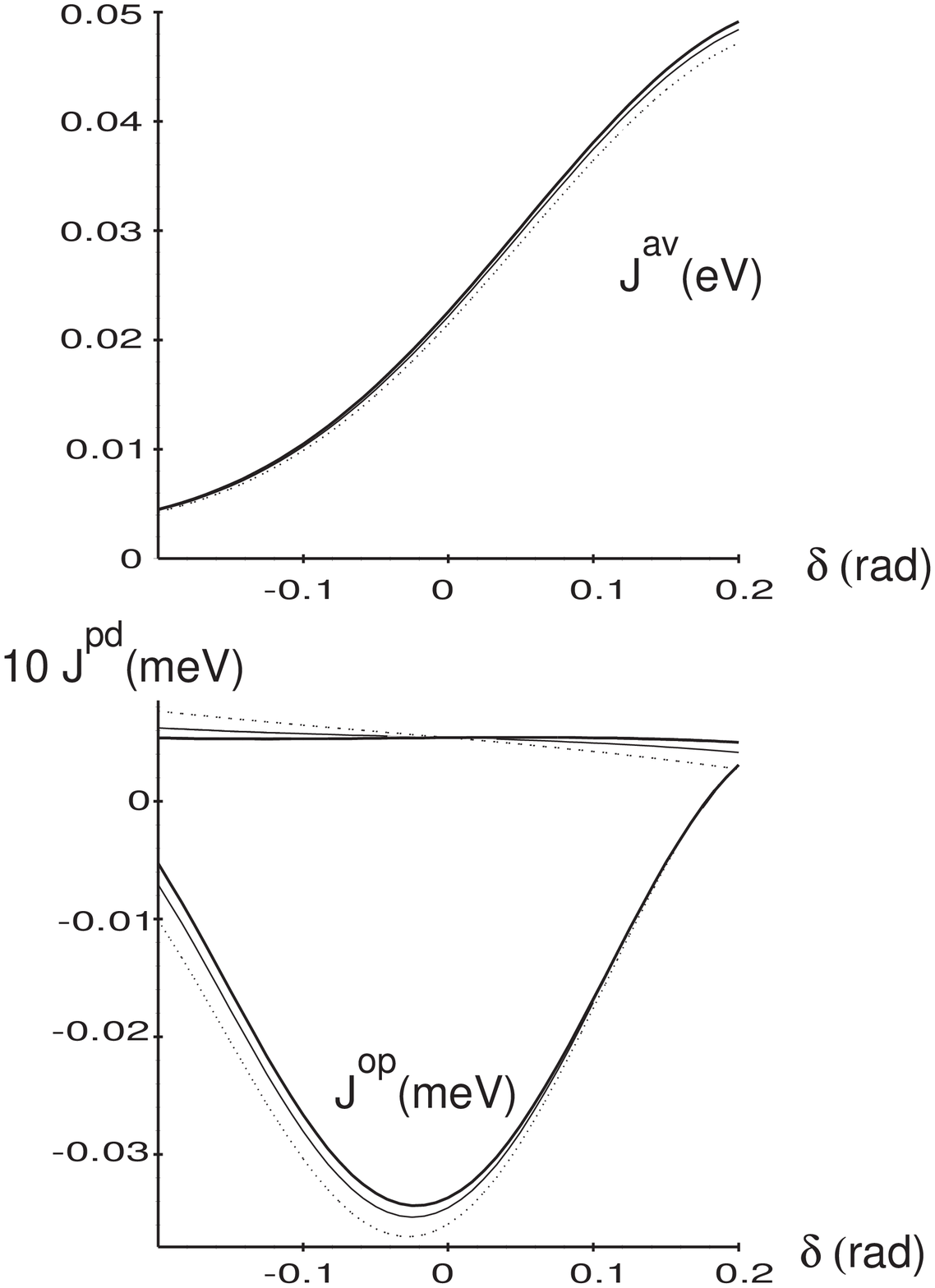}}}
\vspace{2cm}
\caption{ The next--nearest--neighbor magnetic couplings $J_{\rm NNN}$
for K=0.02 eV (bold solid line), K=0.05 (solid line) and
K=0.1 eV (dashed line).}
\label{fig3}
\end{figure}

The parameters that
determine the magnitude and the sign of the magnetic couplings are the
Cu--O and O--O hopping matrix elements, the on--site (single particle)
energies on the oxygen and on the copper, the spin--orbit coupling constant
$\lambda $ and the various Coulomb matrix elements. The latter are parametrized
\cite{griffith}
in terms of the Racah parameters into the on--site leading order interactions,
and the residual remaining interactions, which are small. However, they, as well
as $\lambda$, are necessary for the generation of 
the magnetic anisotropies. 
All these parameters depend on the crystal symmetry,
and hence on the angle $\delta $ (see Fig. \ref{fig1}).
Adopting the plausible assumption that the most important sensitivity
to small deviations from 90$^{\circ}$ occurs in the
Cu--O hopping matrix
elements,
we have included only their dependence on the 
angle, \cite{slater} and calculated the angular dependence of 
the magnetic interactions for small angles $\delta $. The explicit
expressions for the magnetic couplings are given in Secs. III and IV.
Here we summarize the results, which are depicted in Fig. \ref{fig2}
 for the nearest--neighbor couplings, and in Fig. \ref{fig3} for 
the next--nearest--neighbor ones.

The numerical estimates are computed using the following 
parameters. We take the on--site energies on the oxygen to be
$\epsilon_{p_{x}}=\epsilon_{p_{y}}$=3eV and $\epsilon_{p_{z}}$=2eV.
\cite{stein}
Those on the copper are assumed for simplicity to be identical,
$\epsilon_{\alpha}$=1.8eV ($\alpha =0,1,x,y,z$, see definitions below). 
\cite{yildirim} The spin--orbit
coupling on the copper is taken to be $\lambda $=0.1eV. The on--site
Coulomb matrix elements necessitate the Racah parameters $A$, $B$, and
$C$ for the copper, and $F_{0}$ and $F_{2}$ on the oxygen. 
\cite{yildirim,stein} These are chosen as $A$=7.0eV, $B$=0.15eV,
$C$=0.58eV, $F_{0}$=3.1eV, and $F_{2}$=0.28eV. There is no
reliable estimate for the non--local Coulomb exchange
between the copper and the oxygen. \cite{geertsma,braden,mizuno}
We therefore take as a 
representative estimate a value in
the range between 0.02eV and 0.1eV. \cite{geertsma,braden,mizuno} 

The hopping matrix elements can be expressed in terms of the Slater--Koster
parameters, \cite{slater} $t_{0}=-\sqrt{3}(pd\sigma)/2$, 
$t_{1}=-(pd\sigma)/2$,
and $t_{2}=(pd\pi)$, for the Cu--O ones, and 
$t_{3}=(1/2)((pp\sigma)+(pp\pi))$,
$t_{4}=(1/2)((pp\sigma)-(pp\pi))$, and $t_{5}=(pp\pi)$ for the
O--O matrix elements. We have used the values $(pd\sigma)=1.5$eV
\cite{yildirim,stein}
and $(pp\pi)=-0.6$eV \cite{eder}, and used the relations $(pp\pi)
=-\frac{1}{4}(pp\sigma)$ and $(pd\pi)
=-\frac{1}{2}(pd\sigma)$ \cite{matheiss}.

Figure \ref{fig2} depicts the angular dependence of 
$J_{\rm NN}^{\rm av}$, $J_{\rm NN}^{\rm op}$, and $J_{\rm NN}^{\rm pd}$.
The three curves in each figure are obtained by choosing different
representative values for the non--local Cu--O Coulomb 
exchange matrix element, $K$.
At strictly 90$^{\circ}$, $J_{\rm NN}^{\rm av}$ is negative
(ferromagnetic) and small, $\approx -0.02$eV (--0.04eV, --0.07eV)
for $K=0.02$eV (0.05eV, 0.1eV). 
Its value is determined
mainly by the residual Coulomb interactions. As the angle
deviates from 90$^{\circ}$, $J_{\rm NN}^{\rm av}$ approaches
zero and changes its sign at about $\delta\approx $0.05 (0.065, 0.09) 
(2.9$^{\circ}$, 3.7$^{\circ}$, 5.2$^{\circ}$).
These results agree with those found before in Refs. \onlinecite{braden}
and \onlinecite{mizuno}. The out-of--plane anisotropy $J_{\rm NN}^{\rm op}$
is relatively large, and negative, in agreement with the
findings of Ref. \onlinecite{yushankhai}. Like all other anisotropies, its 
magnitude is proportional to $\lambda^{2}$. However, for delicate
reasons related to ``ring--exchange" processes (see below), the
Coulomb matrix element that scales its magnitude is the on--site interaction
on the oxygen, leading to its comparatively high value,
$\approx -1.3$meV at 90$^{\circ}$. (The out--of--plane
anisotropy increases slightly with increasing $K$.)
The nearest--neighbor in--plane anisotropy $J_{\rm NN}^{\rm pd}$
is scaled by the small residual Coulomb interactions. It
vanishes at $\delta =0$, and stays quite small away from that value,
varying approximately linearly with $\delta $, $\approx$ (--0.1 to --1.7 
$K) \delta $ meV.

The case of an ideal 90$^{\circ}$ nearest--neighbor bond has
been recently discussed in Ref. \onlinecite{yushankhai}. These authors
have specialized to materials of the type A$_{2}$Cu$_{3}$O$_{4}$Cl$_{2}$,
with A=Ba or Sr. They have neglected the non--constant
on--site Coulomb interactions
on the oxygen and the non--local Cu--O Coulomb interaction,  but
have taken into account the local orthorhombic symmetry, by allowing 
the Cu on--site energies $\epsilon_{x}$ and $\epsilon_{y}$ to be
different. They therefore obtained a small in--plane anisotropy, 
of the order of 0.2$\mu$eV.

The analogous results for the next--nearest--neighbor
couplings are summarized in Fig. \ref{fig3}. These
necessitate additional perturbation processes, which involve
hopping between nearest--neighbor oxygens. We find that
$J_{\rm NNN}^{\rm av}$ is about 20meV at 90$^{\circ}$,
and has a smooth linear dependence on $\delta $ away from it,
remaining antiferromagnetic for small angles, in agreement with
the findings of Refs. \onlinecite{braden} and \onlinecite{mizuno}.

As in the case of the nearest neighbors, also the anisotropic
coupling $J_{\rm NNN}^{\rm op}$
is relatively large and negative, being $\approx -0.036$meV at
90$^{\circ}$, while $J_{\rm NNN}^{\rm pd}$ is extremely
minute, $\approx -6\mu$eV.
The out--of--plane anisotropy
is again dominated by the ``ring--exchange" processes and the in--plane
anisotropy by the small residual Coulomb interactions $\Delta U$.

Obviously, the results summarized in Figs. \ref{fig2} and \ref{fig3} 
depend on the details of the parameters, e.g., the hopping matrix
elements or the Coulomb Racah coefficients. The remaining Sections
of the paper are devoted to a detailed discussion of the derivation
and the choice of these parameters.

\section{The magnetic Hamiltonian}

As is discussed above, the magnetic Hamiltonian is derived from
a microscopic Hamiltonian. The latter can be written as folllows:
\begin{equation}
{\cal H}={\cal H}_{Cu}+{\cal H}_{O}+{\cal H}_{Cu-O},
\end{equation}
with obvious notations. Explicitly, the Cu--ion Hamiltonian is
\begin{eqnarray}
{\cal H}_{Cu}&=&\sum_{i\alpha\sigma}\epsilon_{\alpha}
d_{i\alpha\sigma}^{\dagger}
d_{i\alpha\sigma}+\frac{\lambda}{2}
\sum_{\stackrel{i\alpha\beta}{\sigma\sigma '}}
{\bf L}_{\alpha\beta}\cdot [\sigma]_{\sigma\sigma '}
d_{i\alpha\sigma}^{\dagger}d_{i\beta\sigma '}\nonumber\\
&+&\frac{1}{2}
\sum_{\stackrel{i\sigma\sigma '}{\alpha\beta\gamma\delta}}
U_{\alpha\beta\gamma\delta}
d_{i\alpha\sigma}^{\dagger}d_{i\beta\sigma '}^{\dagger}
d_{i\gamma\sigma '}d_{i\delta\sigma},\label{HCu}
\end{eqnarray}
where $d_{i\alpha\sigma}^{\dagger}$ creates a hole with spin $\sigma $
in the crystal--field state $\alpha $ at site $i$, 
of site energy $\epsilon_{\alpha}$.
For tetragonal symmetry we label the crystal--field states as
$|0\rangle \sim x^{2}-y^{2}$, $|1\rangle \sim 3z^{2}-r^{2}$,
$|z\rangle\sim xy$, $|x\rangle\sim yz$, and $|y\rangle\sim zx$, where the
$z$ axis is perpendicular to the plane and $|0\rangle$ is the lowest energy
single--particle state. The second term in (\ref{HCu}) is the spin--orbit
interaction, where $\lambda $ is the spin--orbit coupling constant and
${\bf L}_{\alpha\beta}$ denotes the 
matrix elements of the orbital angular
momentum vector between the crystal--field states $\alpha $ and $\beta $.
The non--zero matrix elements are
$L_{0z}^{z}=-2i$, $L_{0x}^{x}=L_{0y}^{y}=i$,
$L_{1x}^{x}=-L_{1y}^{y}=\sqrt{3}i$, $L_{zx}^{y}=-L_{zy}^{x}=L_{xy}^{z}=i$,
and $L_{\alpha\beta}^{\ast}=L_{\beta\alpha}$.
The last term in (\ref{HCu}) is the Coulomb interaction, 
with $U_{\alpha\beta\gamma\delta}=\langle\alpha\delta|\beta\gamma\rangle $
in the notations of Table A26 in Ref. \onlinecite{griffith}. 
The Hamiltonian of the oxygen ions is
\begin{eqnarray}
{\cal H}_{O}&=&\sum_{qn\sigma}\epsilon_{n}p_{qn\sigma}^{\dagger}
p_{qn\sigma}+\sum_{\stackrel{mn\sigma}{qq'}}
(t_{nm}^{qq'}p_{qn\sigma}^{\dagger}p_{q'm\sigma}+hc)\nonumber\\
&+&\frac{1}{2}\sum_{\stackrel{q\sigma\sigma '}{n_{1}n_{2}n_{3}n_{4}}}
U_{n_{1}n_{2}n_{3}n_{4}}p_{qn_{1}\sigma}^{\dagger}
p_{qn_{2}\sigma '}^{\dagger}p_{qn_{3}\sigma '}
p_{qn_{4}\sigma},\label{HO}
\end{eqnarray}
in which $p_{qn\sigma}^{\dagger}$ creates a hole in one of the three
$p$ orbitals, $p_{x}$, $p_{y}$, and $p_{z}$ 
(denoted by $n$) on the oxygen at site $q$, with
energy $\epsilon_{n}$. The second term in (\ref{HO}) describes the
hopping between the O--ions, and the last term is the
Coulomb interaction on the O--ions. 
Finally, ${\cal H}_{Cu-O}$    
describes the kinetic energy of hopping between the Cu and
the O ions, and the Coulomb exchange interaction between them,
\begin{eqnarray}
{\cal H}_{Cu-O}&=&\sum_{\stackrel{iq\sigma}{\alpha n}}(t_{n\alpha}^{qi}
p_{qn\sigma}^{\dagger}d_{i\alpha\sigma}+hc)\nonumber\\
&+&\sum_{\stackrel{iq\sigma\sigma '}{\alpha\beta mn}}
K_{\alpha m\beta n}d_{i\alpha\sigma}^{\dagger}
p_{qm\sigma '}^{\dagger}d_{i\beta\sigma '}p_{qn\sigma}.\label{HCuO}
\end{eqnarray}

A significant simplification of the perturbation expansion is achieved
by first treating the spin--orbit interactions exactly, leaving the
expansion in orders of the spin--orbit coupling, $\lambda $, to
the final stage.\cite{yildirim} This is accomplished by introducing 
the unitary transformation which diagonalizes the single--particle part of
${\cal H}_{Cu}$
\begin{equation}
d_{i\alpha\sigma}^{\dagger}=\sum_{a\sigma '}
[m_{\alpha a}]_{\sigma\sigma '}^{\ast}c_{ia\sigma '}^{\dagger},
\label{trans}
\end{equation}
where $c_{ia\sigma}^{\dagger}$ creates a hole in the exact eigenstate
$a$ of the Hamiltonian which consists of the crystal--field and 
the spin--orbit interaction on the copper. These states have a
site label $i$, a state label $a$, and a pseudo--spin index $\sigma $. 
One then has
\begin{eqnarray}
\sum_{i\alpha\sigma}\epsilon_{\alpha}
d_{i\alpha\sigma}^{\dagger}
d_{i\alpha\sigma}&+&\frac{\lambda}{2}
\sum_{\stackrel{i\alpha\beta}{\sigma\sigma '}}
{\bf L}_{\alpha\beta}\cdot[\sigma]_{\sigma\sigma '}
d_{i\alpha\sigma}^{\dagger}d_{i\beta\sigma '}\nonumber\\
&=&\sum_{ia\sigma}
E_{a}c_{ia\sigma}^{\dagger}c_{ia\sigma},
\end{eqnarray}
where $E_{a}$ and $[m_{\alpha a}]_{\sigma\sigma '}$
are determined by
\begin{eqnarray}
E_{b}[m_{\gamma b}]_{\sigma\sigma '}&=&
\epsilon_{\gamma}[m_{\gamma b}]_{\sigma\sigma '}\nonumber\\
&+&\frac{\lambda}{2}\sum_{\beta\sigma_{1}}
{\bf L}_{\gamma\beta}\cdot[\sigma]_{\sigma\sigma_{1}}
[m_{\beta b}]_{\sigma_{1}\sigma '},
\label{trans1}
\end{eqnarray}
with
$[\sum_{a}{\bf m}_{\beta a}({\bf m}_{\alpha a})^{\dagger}]_{\sigma\sigma '}
=\delta_{\alpha\beta}\delta_{\sigma\sigma '}$. When $\lambda \rightarrow 0$,
each state $|a\rangle $ approaches one of the states $|\alpha\rangle $.
Using this definition, the index $a$ runs over the values 
0, 1, $z$, $x$, and $y$. A detailed discussion
of this transformation is given in Ref. \onlinecite{yildirim}.

To apply the pertubation expansion, 
we divide the Hamiltonian ${\cal H}$ into an 
unperturbed part, ${\cal H}_{0}$, and a perturbation term
${\cal H}_{1}$. The part ${\cal H}_{0}$ contains the single--particle
Hamiltonians on the coppers and on the oxygens, and the leading on--site
Coulomb potentials. The perturbation Hamiltonian contains the kinetic
energy and the residual Coulomb interactions.
As is known, \cite{griffith} the on--site Coulomb interactions
can be parametrized in terms of the Racah coefficients. In
tetragonal site symmetry, those on the copper are parametrized
by the Racah parameters $A$, $B$, and $C$, with $A\gg B$ and $A\gg C$,
and those on the oxygen by $F_{0}$ and $F_{2}$, with $F_{0}\gg F_{2}$.
We choose the on--site leading Coulomb interactions to be $U_{0}\equiv
U_{\alpha\alpha\alpha\alpha}=A+4B+3C$ for the copper, and $U_{q}
\equiv U_{nnnn}=F_{0}+4F_{2}$ on the oxygen. Consequently,
the unperturbed Hamiltonian is
\begin{eqnarray}
{\cal H}_{0}&=&\sum_{ia\sigma}E_{a}c_{ia\sigma}^{\dagger}
c_{ia\sigma}+\frac{U_{0}}{2}\sum_{\stackrel{iab}{\sigma\sigma '}}
c_{ia\sigma}^{\dagger}c_{ib\sigma '}^{\dagger}
c_{ib\sigma '}c_{ia\sigma}\nonumber\\
&+&\sum_{qn\sigma}\epsilon_{n}p_{qn\sigma}^{\dagger}
p_{qn\sigma}+\frac{U_{q}}{2}\sum_{\stackrel{qnn'}{\sigma\sigma '}}
p_{qn\sigma}^{\dagger}p_{qn'\sigma '}^{\dagger}
p_{qn'\sigma '}p_{qn\sigma}.\label{H0}
\end{eqnarray}
The perturbation Hamiltonian is
\begin{equation}
{\cal H}_{1}={\cal H}_{hop}+\Delta{\cal H}_{C},
\end{equation}
in which the hopping term is
\begin{eqnarray}
{\cal H}_{hop}&=&\sum_{\stackrel{iqan}{\sigma\sigma '}}
([\tilde{t}_{an}^{iq}]_{\sigma '\sigma}c_{ia\sigma '}^{\dagger}
p_{qn\sigma}+hc)\nonumber\\
&+&\sum_{\stackrel{nm\sigma}{qq'}}
(t_{nm}^{qq'}p_{qn\sigma}^{\dagger}p_{q'm\sigma}+hc)
.\label{Hhop}
\end{eqnarray}
Because of the transformation (\ref{trans}), the Cu--O
hopping becomes spin--dependent
\begin{equation}
[\tilde{t}_{an}^{iq}]_{\sigma '\sigma}=\sum_{\alpha}
t_{\alpha n}^{iq}[m_{\alpha a}]_{\sigma \sigma '}^{\ast}.\label{ttilde}
\end{equation}
The term $\Delta{\cal H}_{C}$ contains
the (small) additional on--site Coulomb interactions, and the
non--local Cu--O Coulomb potential,
\begin{eqnarray}
\Delta{\cal H}_{C}&=&\frac{1}{2}\sum_{\stackrel{iabcd}{s_{1}s_{2}s_{3}s_{4}}}
\Delta U_{s_{1}s_{2}s_{3}s_{4}}(abcd)c_{ias_{1}}^{\dagger}
c_{ibs_{2}}^{\dagger}c_{ics_{3}}c_{ids_{4}}\nonumber\\
&+&\frac{1}{2}
\sum_{\stackrel{q\sigma\sigma '}{n_{1}n_{2}n_{3}n_{4}}}
\Delta U_{n_{1}n_{2}n_{3}n_{4}}
p_{qn_{1}\sigma}^{\dagger}p_{qn_{2}\sigma '}^{\dagger}
p_{qn_{3}\sigma '}p_{qn_{4}\sigma}\nonumber\\
&+&\sum_{\stackrel{iqabmn}{ss'\sigma\sigma '}}
K_{s\sigma 's'\sigma}(ambn)c_{ias}^{\dagger}
p_{qm\sigma '}^{\dagger}c_{ibs'}p_{qn\sigma},\label{DHC}
\end{eqnarray}
with
\begin{eqnarray}
& &\Delta
U_{s_{1}s_{2}s_{3}s_{4}}(abcd)\nonumber\\
&=&\sum_{\alpha\beta\gamma\delta}
\Delta U_{\alpha\beta\gamma\delta}
[({\bf m}_{\alpha a})^{\dagger}{\bf m}_{\delta d}]_{s_{1}s_{4}}
[({\bf m}_{\beta b})^{\dagger}{\bf m}_{\gamma c}]_{s_{2}s_{3}},
\end{eqnarray}
and
\begin{equation}
K_{s\sigma 's'\sigma}(ambn)=
\sum_{\alpha\beta}K_{\alpha m\beta n}
[m_{\alpha a}]_{\sigma s}^{\ast}[m_{\beta b}]_{\sigma 's'}.
\label{K}
\end{equation}
Here we have defined 
$\Delta U_{\alpha\beta\beta\alpha}=U_{\alpha\beta\beta\alpha}-U_{0}$ 
and $\Delta U_{nn'n'n}=U_{nn'n'n}-U_{q}$. 
For $\alpha\neq\delta $
or $\beta\neq\gamma $, and $n_{1}\neq n_{4}$ or
or $n_{2}\neq n_{3}$,
$\Delta U \equiv U$ involves only the small Racah
coefficients $B$, $C$, and $F_{2}$.\cite{griffith}

\begin{figure}
\epsfxsize=6cm
\centerline{{\epsffile{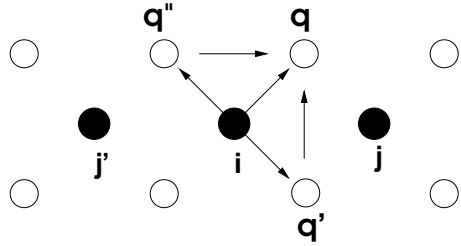}}}
\vspace{1cm}
\caption{The Cu--O effective nearest--neighbor hopping in the
90$^{\circ}$ bond configuration, allowing
for nearest--neighbor O--O hopping. Open circles denote oxygens and
black circles are the Cu's. The processes contributing to
${\tilde t}^{iq}$ are shown by arrows.}
\label{fig4}
\end{figure}

All the perturbation contributions
resulting from ${\cal H}_{1}$
begin and end within the 2$^{\rm N}$--fold degenerate ground--state
manifold of ${\cal H}_{0}$, each state of which has one hole at
each copper site, with arbitrary spin $\sigma $. We will
denote by ``0" the ground state of the single--particle
Hamiltonian on the copper, and will take its site energy to
be zero.
For the sake of clarity, we divide the perturbation channels into
three groups. Group ``a" includes the processes in which there
are two holes on the copper in the intermediate step; group ``b"
includes those in which there are two holes on the oxygen in the
intermediate step, or processes where the two holes exchange their
corresponding coppers by going around the ring formed by the
two coppers and the intervening oxygen (for example, coppers $i$ and
$j$, and oxygens q and $q'$ in Fig. \ref{fig4}); group ``c" contains the
contribution from the non--local Coulomb exchange between the coppers 
and the oxygens. \cite{stechel} 

The contribution of channel ``a" to the magnetic coupling of coppers
$i$ and $j$ is
\begin{eqnarray}
&&{\cal H}_{{\rm a}}(i,j)=\frac{1}{U_{0}}
Tr\Bigl\{{\bf \sigma}\cdot{\bf S}_{i} \tilde{\cal T}_{00}^{ij}
{\bf \sigma}\cdot{\bf S}_{j} \tilde{\cal T}_{00}^{ji}\Bigr\}\nonumber\\
&+&\sum_{\stackrel{ab}{\alpha\beta\gamma\delta}}
\frac{\Delta U_{\alpha\beta\gamma\delta}}
{(U_{0}+E_{a})(U_{0}+E_{b})}\times\nonumber\\
& &\Bigl (Tr\Bigl\{{\bf \sigma}\cdot{\bf S}_{i} \tilde{\cal T}_{0b}^{ij}
({\bf m}_{\alpha b})^{\dagger}{\bf m}_{\delta a}
\tilde{\cal T}_{a0}^{ji}\Bigr\}Tr\Bigl\{{\bf \sigma}\cdot{\bf S}_{j}
({\bf m}_{\beta 0})^{\dagger}{\bf m}_{\gamma 0}\Bigr\}\nonumber\\
& &-Tr\Bigl\{{\bf \sigma}\cdot{\bf S}_{i} \tilde{\cal T}_{0b}^{ij}
({\bf m}_{\alpha b})^{\dagger}
{\bf m}_{\delta 0}{\bf \sigma}\cdot{\bf S}_{j}
({\bf m}_{\beta 0})^{\dagger}{\bf m}_{\gamma a}
\tilde{\cal T}_{a0}^{ji}\Bigr\}\Bigr )\nonumber\\
&+&(i\leftrightarrow j),\label{Ha}
\end{eqnarray}
in which ${\bf S}_{i}$ is the spin on the copper at site $i$
in the new orbital ground state $|a \rangle=|0\rangle $,
\begin{equation}
{\bf S}_{i}=\frac{1}{2}\sum_{\sigma\sigma '}c_{i0\sigma}^{\dagger}
[{\bf \sigma}]_{\sigma\sigma '}c_{i0\sigma '},
\end{equation}
and the ``$Tr$'' are carried out in spin space. \cite{com}
We have introduced in ({\ref{Ha}) the notation 
$\tilde{\cal T}_{ba}^{ij}$ for the effective matrix element for hopping
from state $a$ on copper $j$ to state $b$ on copper $i$. These are
different in the case where the two coppers are nearest
neighbors, and when they are next--nearest neighbors.
In the first case such a process can be achieved through a 
single oxygen, yielding
\begin{equation}
[\tilde{\cal T}_{ba}^{ij}]_{\sigma '\sigma}=\sum_{qn\sigma_{1}}
\frac{1}{\epsilon_{n}}[\tilde{t}_{bn}^{iq}]_{\sigma '\sigma_{1}}
[\tilde{t}_{na}^{qj}]_{\sigma_{1}\sigma},\label{tij}
\end{equation}
to lowest possible order in perturbation theory.
Figure \ref{fig4} depicts the direct hopping
from copper i to oxygen $q$, together with the two possible indirect
hoppings, going through oxygen $q'$ and oxygen $q"$. To account for
these processes one has to use the following replacement in (\ref{tij})
\begin{equation}
[\tilde{t}_{na}^{qi}]_{\sigma_{1}\sigma}\rightarrow
[\tilde{t}_{na}^{qi}]_{\sigma_{1}\sigma}-
\sum_{mq'}\frac{1}{\epsilon_{m}}t_{nm}^{qq'}
[\tilde{t}_{ma}^{q'i}]_{\sigma_{1}\sigma}.\label{rep}
\end{equation}
When the two coppers are NNN, the O--O hopping is essential 
for bringing the two holes to the same copper. (For example, the bond
$j'$--$j$ in Fig. \ref{fig4} requires the $q$--$q"$ hopping.)
In this case we have
\begin{equation}
[\tilde{\cal T}_{ba}^{ij}]_{\sigma '\sigma}=
\sum_{qn\sigma_{1}}\frac{1}{\epsilon_{n}}
[\tilde{t}_{bn}^{iq}]_{\sigma '\sigma_{1}}
[\tilde{T}_{na}^{qi}]_{\sigma_{1}\sigma},
\end{equation}
where
\begin{equation}
[\tilde{T}_{na}^{qi}]_{\sigma_{1}\sigma}=
\sum_{mq'}\frac{1}{\epsilon_{m}}t_{nm}^{qq'}
[\tilde{t}_{ma}^{q'i}]_{\sigma_{1}\sigma}.\label{T}
\end{equation}

The perturbation contributions coming from 
channel b, when coppers $i$ and $j$ are NN, yield
\begin{eqnarray}
&&{\cal H}_{{\rm b}}(i,j)=\sum_{mnqq'}
\Bigl (\frac{1}{\epsilon_{n}}+\frac{1}{\epsilon_{m}}
\Bigr )^{2}\times\nonumber\\
&&\Bigl (\delta_{qq'}\frac{1}{\epsilon_{n}+\epsilon_{m}+U_{q}}
+(1-\delta_{qq'})\frac{1}{\epsilon_{n}+\epsilon_{m}}
\Bigr )\times\nonumber\\
&&Tr\Bigl\{{\bf \sigma}\cdot{\bf S}_{i}\tilde{t}_{0m}^{iq'}
\tilde{t}_{m0}^{q'j}{\bf \sigma}\cdot{\bf S}_{j}\tilde{t}_{0n}^{jq}
\tilde{t}_{no}^{qi}\Bigr\}\nonumber\\
&&+
\sum_{\stackrel{qnm}{n'm'}}\Bigl
(\frac{1}{\epsilon_{n}}+\frac{1}{\epsilon_{n'}}
\Bigr )\Bigl (\frac{1}{\epsilon_{m}}+\frac{1}{\epsilon_{m'}}
\Bigr )\times\nonumber\\
& &\frac{\Delta U_{mm'nn'}}{(U_{q}+\epsilon_{n}
+\epsilon_{n'})(U_{q}+\epsilon_{m}+\epsilon_{m'})}
\times\nonumber\\
&&\Bigl (Tr\Bigl\{{\bf \sigma}\cdot{\bf S}_{i}\tilde{t}_{0m}^{iq}
\tilde{t}_{n'0}^{qi}\Bigr\} Tr\Bigl\{{\bf \sigma}\cdot{\bf S}_{j}
\tilde{t}_{0m'}^{jq}\tilde{t}_{n0}^{qj}\Bigr\}\nonumber\\
&&-Tr\Bigl\{{\bf \sigma}\cdot{\bf S}_{i}\tilde{t}_{0m}^{iq}
\tilde{t}_{n'0}^{qj}{\bf \sigma}\cdot{\bf S}_{j}\tilde{t}_{0m'}^{jq}
\tilde{t}_{n0}^{qi}\Bigr\}\Bigr ).\label{Hb}
\end{eqnarray}
The term with the $(1-\delta_{qq'})$ in front
arises in the chain geometry and does not have the on--site Coulomb
interaction in the denominator.\cite{geertsma,eskes} As non--local
Coulomb interactions between the oxygens are ignored here, there
is no analogous contribution in the second sum of (\ref{Hb}).
When the two coppers are NNN, one must invoke the
O--O hopping. We then find that each pair of matrix elements
$\tilde{t}$ has to be replaced as follows
\begin{equation}
\tilde{t}^{qi}\tilde{t}^{qj}\rightarrow 
-\tilde{T}^{qi}\tilde{t}^{qj}-\tilde{t}^{qi}\tilde{T}^{qj},\label{repb}
\end{equation}
where both $\tilde{t}$ and $\tilde{T}$ are matrices in spin space,
given by Eqs. (\ref{ttilde}) and (\ref{T}), respectively.

Finally, channel c gives
\begin{eqnarray}
&&{\cal H}_{{\rm c}}(i,j)=\nonumber\\
&&-\sum_{\stackrel{qmn}{\alpha\gamma}}
\frac{K_{\alpha m\gamma n}}{\epsilon_{n}\epsilon_{m}}
Tr\Bigl\{{\bf \sigma}\cdot{\bf S}_{i}(
{\bf m}_{\alpha 0})^{\dagger}
\tilde{t}_{n0}^{qj}{\bf \sigma}\cdot{\bf S}_{j}
\tilde{t}_{0m}^{jq}{\bf m}_{\gamma 0}\Bigr\}\nonumber\\
&&+(i\leftrightarrow j),\label{Hc}
\end{eqnarray}
when the two coppers are NN. 
The corresponding expression for NNN coppers is obtained from
(\ref{Hc}) by replacing each matrix element $\tilde{t}$ by $\tilde{T}$.

The total effective magnetic interaction is the sum of 
the three groups. This is now expanded up to second order in the
spin--orbit coupling $\lambda $. This requires the
matrix $m$, Eq. (\ref{trans}), up to first order in $\lambda$ only. 
Using Eq. (\ref{trans1}) we have
\begin{equation}
[m_{\alpha a}]_{\sigma\sigma '}=\delta_{\alpha a}\delta_{\sigma\sigma '}
-\frac{\lambda}{2}\frac{{\bf L}_{\alpha a}\cdot[\sigma]_{\sigma\sigma '}}
{\epsilon_{\alpha}-E_{a}}.
\end{equation}
A further significant simplification of the 
expressions is achieved when one takes into
account the symmetry properties of the Coulomb matrix 
elements:\cite{yildirim,stein} $\Delta U_{\alpha\beta\gamma\delta}$
vanishes unless the values of $\alpha $, $\beta $, $\gamma $, 
and $\delta $ are such that the products $\sigma (\alpha )\sigma (\delta )$
and $\sigma (\beta )\sigma (\gamma )$ are proportional to
each other. Here we use the convention that 
$\sigma (\alpha =0)$ and $\sigma (\alpha =1)$ are the unit
matrix. Similarly, the  non--vanishing matrix elements of 
$\Delta U_{n_{1}n_{2}n_{3}n_{4}}$ satisfy $\sigma (n_{1})\sigma (n_{4})\propto
\sigma (n_{2})\sigma (n_{3})$, and those of
$K_{\alpha m\beta n}$
vanish unless $\sigma (\alpha )\sigma (n)\propto \sigma (\beta )\sigma (m)$.

In the following we present the general expressions in the form
\begin{equation}
{\cal H}(i,j)=\left(J-\frac{1}{2} {\rm Tr} {\bf \Gamma} \right)
{\bf S}_{i}\cdot{\bf S}_{j}+{\bf S}_{i}{\bf \Gamma}{\bf S}_{j},
\end{equation}
where $J$ includes all contributions to zeroth order in $\lambda $,  
while the matrix ${\bf \Gamma}$ 
contains the contributions which necessitate the spin--orbit interaction,
and is
therefore second--order in $\lambda $.
The application of these general expressions 
to the specific Cu--Cu bonds will be carried out in the next Section.

(a) Channel a [Eq. (\ref{Ha})] yields
\begin{equation}
J_{{\rm a}}=4\Bigl (\frac{t_{00}^{ij}t_{00}^{ji}}{U_{0}}
-\sum_{\alpha}\frac{\Delta U_{\alpha 0\alpha 0}
t_{0\alpha}^{ij}t_{\alpha 0}^{ji}}{(U_{0}+\epsilon_{\alpha})^{2}}
\Bigr ),\label{Ja}
\end{equation}
and
\begin{eqnarray}
&&\Gamma_{{\rm a}}^{\mu\nu}=
\frac{\lambda^{2}}{2U_{0}}\frac{L_{\mu 0}^{\mu}L_{\nu 0}^{\nu}}
{\epsilon_{\mu}\epsilon_{\nu}}\times\nonumber\\
&&\Biggl (\Bigl [ \bigl (t_{0\mu}^{ij}-t_{\mu 0}^{ij}
\bigr )\bigl (t_{0\nu}^{ji}-t_{\nu 0}^{ji}\bigr )
+\mu\leftrightarrow\nu\Bigr ]+j\leftrightarrow i\Biggr )\nonumber\\
&-& \frac{\lambda^2}{2}\sum_{\alpha \beta \gamma \delta}\frac{\Delta U_{\alpha \beta \gamma \delta}}
{(U_{0}+\epsilon_{\alpha})(U_{0}+\epsilon_{\gamma})} \times \\ \nonumber
& & \Bigl (\delta_{\delta 0}\sum_{\alpha '}\frac{t_{0\alpha '}^{ij}
L_{\alpha '\alpha}^{\mu}}{U_{0}+\epsilon_{\alpha '}}+
\frac{L_{\mu 0}^{\mu}}{\epsilon_{\mu}}
(t_{0\alpha}^{ij} \delta_{\mu \delta}-t_{\mu\alpha}^{ij}\delta_{\delta 0})
\Bigr )\times\nonumber\\
&&\Bigl (\delta_{\beta 0}\sum_{\alpha '}\frac{t_{\alpha '0}^{ji}
L_{\gamma\alpha '}^{\nu}}{U_{0}+\epsilon_{\alpha '}}+
\frac{L_{\nu 0}^{\nu}}{\epsilon_{\nu}}
(-t_{\gamma 0}^{ji}\delta_{\nu \beta}+t_{\gamma\nu}^{ji}\delta_{\beta 0})
\Bigr )\nonumber\\
&+&(\mu \leftrightarrow \nu + j \leftrightarrow i)  .\label{Ga}
\end{eqnarray}}
(b) Channel b [Eq. (\ref{Hb})] yields
\begin{eqnarray}
&&J_{{\rm b}}=2\sum_{mnqq'}\Bigl
(\frac{1}{\epsilon_{n}}+\frac{1}{\epsilon_{m}}\Bigr )^{2}
t_{0m}^{iq'}t_{m0}^{q'j}t_{0n}^{jq}t_{n0}^{qi}\times\nonumber\\
&&\Bigl (\delta_{qq'}\frac{1}{\epsilon_{n}+\epsilon_{m}+U_{q}}
+(1-\delta_{qq'})\frac{1}{\epsilon_{n}+\epsilon_{m}}\Bigr )\nonumber\\
&-&2\sum_{mnq}
\Bigl (\frac{1}{\epsilon_{n}}+\frac{1}{\epsilon_{m}}\Bigr )^{2}
\frac{\Delta U_{mnmn}(t_{0m}^{iq}t_{n0}^{qj})^{2}}
{(\epsilon_{n}+\epsilon_{m}+U_{q})^{2}}\nonumber\\
&-&8\sum_{nmq}\frac{1}{\epsilon_{n}\epsilon_{m}}
\frac{\Delta
U_{mmnn}t_{0m}^{iq}t_{n0}^{qj}t_{om}^{jq}t_{n0}^{qi}}
{(U_{q}+2\epsilon_{n})(U_{q}+2\epsilon_{m})},\label{Jb}
\end{eqnarray}
and
\begin{eqnarray}
&&\Gamma_{{\rm b}}^{\mu\nu}=\lambda^{2}\frac{L_{\mu 0}^{\mu}}{\epsilon_{\mu}}
\frac{L_{\nu 0}^{\nu}}{\epsilon_{\nu}}
\sum_{mnqq'}\Biggl\{\Bigl (\frac{1}{\epsilon_{n}}+
\frac{1}{\epsilon_{m}}\Bigr )^{2}\times\nonumber\\
&&\Bigl [\Bigl (\delta_{qq'}\frac{1}{\epsilon_{n}+\epsilon_{m}+U_{q}}
+(1-\delta_{qq'})\frac{1}{\epsilon_{n}+\epsilon_{m}}\Bigr )\times\nonumber\\
&&T_{0\mu}^{ij}(m,m,q')T_{0\nu}^{ji}(n,n,q)+
\delta_{qq'}\frac{\Delta U_{mnmn}}{(\epsilon_{n}
+\epsilon_{m}+U_{q})^{2}}\times\nonumber\\
&&\Bigl (T_{0\mu}^{ii}(m,n,q)T_{0\nu}^{jj}(n,m,q)-
T_{0\mu}^{ij}(m,n,q)T_{0\nu}^{ji}(n,m,q)\Bigr )\Bigr ]\nonumber\\
&+&\delta_{qq'}\frac{4}{\epsilon_{n}\epsilon_{m}}
\frac{\Delta U_{mmnn}}{(U_{q}+2\epsilon_{n})(U_{q}
+2\epsilon_{m})}\times\nonumber\\
&&\Bigl [T_{0\mu}^{ii}(m,n,q)T_{0\nu}^{jj}(m,n,q)
-\frac{1}{2}\Bigl (T_{0\mu}^{ij}(m,n,q)T_{0\nu}^{ji}(m,n,q)\nonumber\\
&+&T_{0\nu}^{ij}(m,n,q)T_{0\mu}^{ji}(m,n,q)\Bigr )\Bigr ]\Biggr\},
\label{Gb}
\end{eqnarray}
where we have defined
\begin{equation}
T_{0\mu}^{ij}(m,n,q)=t_{0m}^{iq}t_{n\mu}^{qj}-t_{\mu m}^{iq}
t_{n0}^{qj}.
\end{equation}
(c) Channel c [Eq. (\ref{Hc})] yields
\begin{equation}
J_{{\rm c}}=-2\sum_{nq}K_{0n0n}
\frac{t_{n0}^{qj}t_{0n}^{jq}}{\epsilon_{n}^{2}}+(j\rightarrow i),
\label{Jc}
\end{equation}
and
\begin{eqnarray}
&&\Gamma_{{\rm c}}^{\mu\nu}=-\lambda^{2}\sum_{mnq}
\frac{1}{\epsilon_{m}\epsilon_{n}}\frac{L_{\mu 0}^{\mu}
L_{\nu 0}^{\nu}}{\epsilon_{\mu}\epsilon_{\nu}}\times\nonumber\\
&&\Bigl (K_{\mu m0n}t_{n0}^{qj}t_{\nu m}^{jq}-
K_{\mu m\nu n}t_{n0}^{qj}t_{0m}^{jq}\nonumber\\
&+&
K_{0m\nu n}t_{n\mu}^{qj}t_{0m}^{jq}-
K_{0m0n}t_{n\mu}^{qj}t_{\nu m}^{jq}+(j\rightarrow i)\Bigr ).
\label{Gc}
\end{eqnarray}

For simplicity, we have written the results for $J$
and ${\bf \Gamma}$, Eqs. (\ref{Ja}--\ref{Gc}) for the 
nearest--neighbor Cu--Cu bond. The analogous expressions for 
the next--NN bond are obtained using the replacements (\ref{T})
and (\ref{repb}).

\section{The magnetic couplings}

\subsection{The nearest--neighbor 90$^{\circ}$ bond}

\begin{figure}
\epsfxsize=10cm
\centerline{{\epsffile{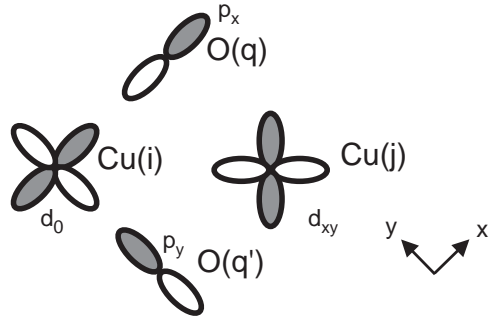}}}
\vspace{-5cm}
\caption{
Cu and O orbitals which are involved in the processes
leading to the out-of-plane anisotropy $\Gamma^{zz}$.
Here and below, the shaded (white) area indicates positive (negative) phase.}
\label{fig5}
\end{figure}

We list in Tables \ref{tab1} and \ref{tab2} the hopping matrix elements
between Cu and O for the 90$^{\circ}$ configuration (see the
Introduction and Fig. \ref{fig4}
for the notations). 
Using these values, we find that the leading order
contributions to the magnetic couplings come from channels ``b" and ``c",
with
\begin{eqnarray}
&J&=J_{{\rm b}}+J_{{\rm c}}=\nonumber\\
&-&\frac{16t_{0}^{4}\Delta U_{p_{x}p_{y}p_{x}p_{y}}}{\epsilon_{p_{x}}^{2}
(2\epsilon_{p_{x}}+U_{q})^{2}}-\frac{8t_{0}^{2}}{\epsilon_{p_{x}}^{2}}
K_{0p_{x}0p_{x}}.
\label{J90}
\end{eqnarray}
in accordance with the GKA rules. Inserting the numerical values of
the parameters, we find
$J \approx -8$meV$-0.67 K_{0p_x0p_x}$
The leading order magnetic anisotropy in this case is the
out--of--plane one,  
\begin{equation}
\Gamma_{{\rm leading}\ {\rm order}}^{zz}=-\frac{64
 \lambda ^{2}t_{0}^{2}t_{2}^{2}U_{q}} 
{\epsilon_{z}^{2}\epsilon_{p_{x}}^{3}(2\epsilon_{p_{x}}+U_{q})},
\label{G90}
\end{equation}
with $\Gamma^{zz} \approx -1.3$meV.
The processes yielding the latter are depicted in Fig. \ref{fig5}.

The remaining small anisotropies resulting
from the on--site Coulomb potential on the oxygen, and from the non--local
Coulomb exchange between the copper and the oxygen, are listed
below. We express those in the
coordinate system depicted in Fig. \ref{fig5}. To obtain the couplings
in the coordinate system discussed in the Introduction, one has
to rotate  by 45$^{\circ}$.
\begin{eqnarray}
&&\Delta\Gamma_{{\rm a}}^{zz}=
32 \lambda^2
\frac{(4B+C)(t_0t_2)^2}{\epsilon_{p_x}^2\epsilon_z^2 
(\epsilon_1+U_{0})^2},
 \nonumber \\
&&\Delta\Gamma_{{\rm a}}^{xx}=\Delta\Gamma_{{\rm a}}^{yy}
=2 \lambda^2
\frac{(3B+C)t_2^4}{\epsilon_{p_z}^2\epsilon_x^2 
(\epsilon_x+U_{0})^2},
\nonumber \\ 
&&\Delta\Gamma^{xx}_{{\rm b}}=\Delta\Gamma^{yy}_{{\rm b}}=
-\frac{\lambda^{2}t_{0}^{2}t_{2}^{2} 3F_2}{\epsilon_{x}^{2}
(\epsilon_{p_{x}}+\epsilon_{p_{z}}+U_{q})^{2}}
\Bigl (\frac{1}{\epsilon_{p_{z}}}+\frac{1}{\epsilon_{p_{x}}}
\Bigr )^{2},\nonumber \\
&&\Delta\Gamma_{{\rm c}}^{zz}=-\frac{8\lambda^{2}}
{\epsilon_{z}^{2}\epsilon_{p_{x}}^{2}}\Bigl (2 t_{0}^{2}K_{zp_{x}zp_{x}}
+2 t_{2}^{2} K_{0p_{x}0p_{x}}  \nonumber\\
&&+4 t_{0}t_{2} K_{zp_{y}0p_{x}}\Bigr ),\nonumber\\
&&\Delta\Gamma_{{\rm c}}^{xx}=\Delta\Gamma_{{\rm c}}^{yy}=-\frac{2\lambda^{2}}
{\epsilon_{y}^{2}}\Bigl (\frac{t_{0}^{2}}{\epsilon_{p_{x}}^{2}}
(K_{yp_{x}yp_{x}}+K_{yp_{y}yp_{y}})\nonumber\\
&+&\frac{t_{2}^{2}}{\epsilon_{p_{z}}^{2}}
K_{0p_{z}0p_{z}}\Bigr ).
\end{eqnarray}
To obtain these results we
have used the relations \cite{griffith}
$U_{1010}=4B+C$, $U_{x0x0}=U_{y0y0}=3B+C$,
$\Delta U_{p_{z}p_{y}p_{z}p_{y}}=\Delta U_{p_{z}p_{x}p_{z}p_{x}}=3F_2$,
$\Delta U_{p_{x}p_{x}p_{y}p_{y}}=\Delta U_{p_{x}p_{y}p_{x}p_{y}}$,
$K_{z p_y 0 p_x}=-K_{z p_x 0 p_y},K_{x p_z 0 p_y}=K_{y p_z 0 p_x}$,
$K_{0 p_y 0 p_y}=-K_{0 p_x 0 p_x},K_{z p_x z p_x}=K_{z p_y z p_y}$.
In addition, from Tables \ref{tab1} and \ref{tab2}, 
the only non--zero effective Cu--Cu hoppings are
\begin{eqnarray}
t_{z1}^{ij}=t_{z1}^{ij}=\frac{2 t_1 t_2}{\epsilon_{p_x}},  \ \
t_{xy}^{ij}=t_{yx}^{ij}=\frac{2 t_2^2}{\epsilon_{p_z}}.
\end{eqnarray}
Numerical estimates of these expressions yield
$\Delta\Gamma_{{\rm a}}^{zz} \approx 40 \mu$eV$,\Delta\Gamma_{{\rm a}}^{xx} 
\approx 7
\mu$eV,  $\Delta\Gamma_{{\rm b}}^{xx} \approx -9 \mu$eV, 
$\Delta\Gamma_{{\rm c}}^{zz}
\approx -K \times 0.003,$  $\Delta\Gamma_{{\rm c}}^{xx}
\approx -K \times 0.001$.

\subsection{The next--nearest--neighbor 90$^{\circ}$ bond}

\begin{figure}
\epsfxsize=8cm
\centerline{{\epsffile{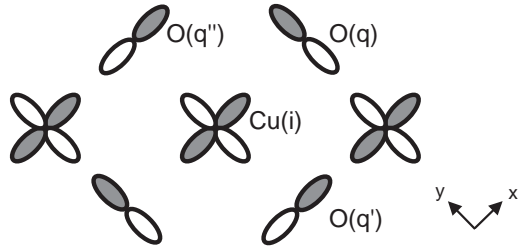}}}
\vspace{-5cm}
\hspace{-1cm}
\caption{
Next--nearest--neighbor hopping between two $d_{x^2-y^2}$
orbitals via the $p_x$ and $p_y$ orbitals.}
\label{fig6}
\end{figure}

Tables \ref{tabn1} and \ref{tabn2} list the hopping matrix elements
$T_{\alpha n}^{iq}$ for a
Cu--O--O--Cu process in the 90$^{\circ}$ configurtion (the notations
are shown in Fig. \ref{fig4}). In these Tables, $t_{3}=t_{p_{x}p_{x}}=
t_{p_{y}p_{y}}$, $t_{4}=t^{qq'}_{p_{x}p_{y}}=-t^{qq"}_{p_{x}p_{y}}$,
and $t_{5}=t_{p_{z}p_{z}}$.
The effective hopping matrix elements between two coppers,
\begin{eqnarray}
{\cal T}_{ab}^{ij}=\sum_{q q' m n} \frac{1}{\epsilon_n \epsilon_m} 
t_{a n}^{iq} t_{nm}^{qq'} t_{mb}^{q'j},
\end{eqnarray}
which do not vanish are
\begin{eqnarray}
&&{\cal T}_{00}^{ij}= \frac{-2 t_0^2 t_4}{\epsilon_{p_x}^2},\ \
{\cal T}_{11}^{ij}= \frac{2 t_1^2 t_4}{\epsilon_{p_x}^2}, \ \
{\cal T}_{zz}^{ij}= \frac{2 t_2^2 t_4}{\epsilon_{p_x}^2}, \nonumber\\
&&{\cal T}_{z1}^{ij}={\cal T}_{1z}^{ij}= 
\frac{4 t_1 t_2 t_3}{\epsilon_{p_x}^2},\ \ 
{\cal T}_{xy}^{ij}={\cal T}_{yx}^{ij}= 
\frac{t_2^2 t_5}{\epsilon_{p_z}^2}.
\end{eqnarray}
As opposed to the NN--bond, 
in this case hopping between the ground state orbitals of NNN coppers
is possible, via the two oxygen orbitals $p_{x}$ and $p_{y}$, 
which are connected by $t_{4}$, see Fig. \ref{fig6}.
We find contributions to the  coupling $J$ 
from all three channels,
\begin{eqnarray}
&& J=J_a+J_b+J_c, \nonumber \\
&&J_a= \frac{64 t_0^{4}t_4^2 }{\epsilon_{p_x}^4 U_0}, \nonumber\\
&&J_b=   
\frac{32 t_0^{4}t_4^2 (U_q+4\epsilon_{p_x})}{\epsilon_{p_x}^5 
(2\epsilon_{p_x}+U_q) } 
-\frac{32 t_0^{4}(t_4^2+t_3^2) \Delta U_{p_xp_yp_xp_y}}
{\epsilon_{p_x}^4 (2\epsilon_{p_x}+U_{q})^2},\nonumber\\
&&J_c=-8\frac{K_{0p_x0p_x}}{\epsilon_{p_x}^4}t_0^2 (t_3^2+t_4^2),
\end{eqnarray}
with
$J \approx 0.02$eV
The leading order anisotropy is the out--of--plane one, and comes 
mainly from channel ``b"
\begin{equation}
\Gamma_{{\rm leading}\ {\rm order}}^{zz}=-\frac{256
 \lambda ^{2}t_{0}^{2}t_{2}^{2}t_3^2 U_{q}} 
{\epsilon_{z}^{2}\epsilon_{p_{x}}^{5}(2\epsilon_{p_{x}}+U_{q})},
\label{GN90}
\end{equation}
$\Gamma_{{\rm leading}\ {\rm order}}^{zz} \approx -30 \mu$eV.
The remaining non--diagonal anisotropies, caluclated in the coordinate system
of Fig. \ref{fig6},  are
\begin{eqnarray}
&&\Delta \Gamma_a^{xy}=-64 \lambda^2 \Delta U_{x0x0} 
\frac{t_0^2t_2^2t_4t_5}{(\epsilon_x+U_0)\epsilon_x^2\epsilon_{p_z}^2
\epsilon_{p_x}^2},\nonumber\\
&&\Delta \Gamma_b^{xy}=-16 \lambda^2 \Delta U_{p_xp_xp_zp_z}\times\nonumber\\
&& 
\frac{t_2^2t_0^2 t_4 \Bigl( t_{3}/\epsilon_{p_{x}}+
t_{5}/\epsilon_{p_{z}}\Bigr)
}{\epsilon_x^2 \epsilon_{p_x}^2\epsilon_{p_z}
(2\epsilon_{p_x}+U_q)(2\epsilon_{p_z}+U_q)} ,\nonumber\\
&&\Delta \Gamma_c^{xy}= -2 \frac{\lambda^2}{\epsilon_x^2\epsilon_{p_x}^4}
t_0^2t_3t_4 (K_{xp_xyp_y}+K_{xp_yyp_x}).
\end{eqnarray}
Here $\Gamma^{xy}=\Gamma^{yx}$. As before, this coordinate system has
to be rotated by 45$^{\circ}$ in order to produce the magnetic couplings
in the form discussed in the Introduction. That is,
\begin{eqnarray}
\Gamma^{xx}+\Gamma^{xy} \rightarrow \Gamma^{xx},\ \
\Gamma^{yy}-\Gamma^{xy}\rightarrow \Gamma^{yy}.
\end{eqnarray}

\section{Small deviation from 90$^{\circ}$}

\begin{figure}
\epsfxsize=10cm
\centerline{{\epsffile{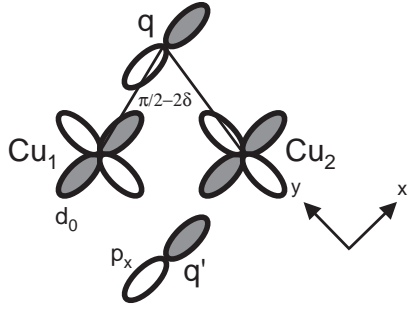}}}
\vspace{-6cm}
\caption{
Cu--O--Cu geometry for an angle $\pi/2-2 \delta$.}
\label{fig7}
\end{figure}

To calculate the magnetic couplings as  function of the angle
$\delta $ (see Figs. \ref{fig1} and \ref{fig7})
we use the following forms for the hopping matrix elements
\begin{eqnarray}
& &t_{0 p_x}^{1q}= t_0 \cos \delta (1-2 \cos^2 \delta) 
+2 t_2 \sin \delta \cos \delta , \nonumber\\
& & t_{0 p_y}^{1q}= -t_0 \sin \delta (1-2 \cos^2 \delta) 
+2 t_2 \sin \delta \cos^2 \delta , 
\end{eqnarray}
with $t_{0 p_x}^{q1}=-t_{0 p_y}^{q2}$ and $t_{0 p_y}^{q1}=-t_{0 p_x}^{q2}$, 
\begin{eqnarray}
& &t_{z p_x}^{1q}=2t_0 \sin \delta \cos^2 \delta+t_2 \sin \delta
(2 \cos^2 \delta-1 ),  \nonumber\\
& &t_{z p_y}^{1q}=-2t_0 \cos \delta \sin^2 \delta+t_2 \cos \delta
(2 \cos^2 \delta -1 ), 
\end{eqnarray}
with $t_{z p_x}^{q1}=t_{z p_y}^{q2}$ and $t_{z p_y}^{q1}=t_{z p_x}^{q2}$,
\begin{eqnarray}
& &t_{1 p_x}^{1q}= t_1 \cos \delta ,\nonumber\\
& &t_{1 p_x}^{1q}= -t_1 \sin \delta ,
\end{eqnarray}
with $t_{1 p_x}^{q1}=t_{1 p_y}^{q2}$ and $t_{1 p_y}^{q1}=t_{1 p_x}^{q2}$,
and
\begin{eqnarray}
& &t_{x p_z}^{1q}= -t_2 \sin \delta ,\nonumber\\
& &t_{y p_z}^{1q}= t_2 \cos \delta ,
\end{eqnarray}
with $t_{x p_z}^{q1}=t_{y p_z}^{q2}$.

For the next--nearest neighbor hopping we need the following
matrix elements:
\begin{eqnarray}
T_{p_x \alpha}=\frac{1}{\epsilon_{p_x}}(t_3 t_{p_x \alpha}
+t_4 t_{p_y \alpha}), \nonumber\\
T_{p_y \alpha}=\frac{1}{\epsilon_{p_x}}(t_3 t_{p_y \alpha}
+t_4 t_{p_x \alpha}), 
\end{eqnarray}
for $\alpha=0,1,z$,
and
\begin{eqnarray}
T_{p_z \alpha}=\frac{1}{\epsilon_{p_z}}t_5 t_{p_z \alpha}
\end{eqnarray}
for $\alpha=x,y$.
The resulting explicit expressions for the magnetic couplings
are very long, and we therefore skip them. Instead, we have used the
results above to produce the curves in Figs. \ref{fig2}
and \ref{fig3}, to obtain the magnetic couplings as function
of the angular deviation $\delta $.

\section{Discussion}

We have presented a detailed calculation of the magnetic
interaction between nearest--neighbor and next--nearest--neighbor
coppers in the edge--sharing geometry, and obtained numerical estimates
for the various couplings as function of the angular deviation from
90$^{\circ}$. These numerical estimates are crucial for the analysis
of the magnetic structures of many chain, ladder and lamellar
cuprates. Our calculation is based on a perturbation expansion of 
a general Hubbard Hamiltonian. It has been found before
that for the magnetic {\it anisotropies} of the linear Cu--O--Cu bond,
this expansion is quite reliable. \cite{yildirim} 
On the other hand, it has been argued, (again for the 180$^{\circ}$--bond),
that perturbation theory fails to yield reasonable values for the leading
magnetic {\it isotropic intractions}, \cite{eskes} because the hopping
matrix elements $t$ are not necessarily small compared with the
on--site energies. However, the almost 90$^{\circ}$ case discussed here
is different, because of the appearance of the small Coulomb
matrix elements in the expansion. It therefore can be expected that
the perturbation expansion for the present case yields reliable
estimates. Indeed, the comparison of our results for the NN
isotropic energy with those obtained from exact diagonalization
\cite{mizuno} seem to support this conclusion.

Our results show that the out--of--plane anisotropy is negative,
both for NN and for NNN coppers. This indicates an easy axis perpendicular
to the Cu--O plane, in agreement with Ref. \onlinecite{yushankhai}. We find
that the
pseudo--dipolar interaction between nearest--neighbors 
vanishes at strictly 90$^{\circ}$, and is
minute for a small deviation away from it. As has been shown in Ref.
\onlinecite{yushankhai}, this result is modified when one
allows for a difference between the Cu on--site
energies $\epsilon_{x}$ and $\epsilon_{y}$. It seems that this should be the
case in materials like Sr$_{2}$Cu$_{3}$O$_{4}$Cl$_{2}$, where some
of the copper ions lose their local tetragonal symmetry: It has been found
\cite{geillo} that experimental data on Sr$_{2}$Cu$_{3}$O$_{4}$Cl$_{2}$
imply a finite value for this energy. This means that the
interpretation of the data necessitates the inclusion of such
effects, or of dipolar
interactions. In the same manner, it is expected
that our numerical estimates as function of the angle $\delta $
will be useful
in the analysis of other cuprates.

\acknowledgements
We have benefitted from discussions with A. B. Harris.
This project has been supported by a grant from the U. S.--Israel 
Binational Science Foundation (BSF).
S. T. acknowledges the support by the Deutsche Forschungsgemeinschaft.

\end{multicols}

\begin{table}
\caption{The Cu--O hopping matrix elements $t_{\alpha n}^{iq}$ {\it for the
90$^{\circ}$ bond}. Upper signs: $t^{iq}$, lower signs: $t^{jq'}$.}
\label{tab1}
\begin{tabular}{c|ccc}
  & $p_{x}$ & $p_{y}$ & $p_{z}$   \\  \hline
0 & $\mp t_{0}$       & 0       & 0         \\
1 & $\pm t_{1}$ &  0      & 0         \\
$z$ & 0 & $\pm t_{2}$ &0 \\
$x$ & 0 & 0 & 0 \\
$y$ & 0 & 0 & $\pm t_{2}$  \\ \hline
\end{tabular}
\end{table}
\begin{table}
\caption{The Cu--O hopping matrix elements $t_{\alpha n}^{iq}$ {\it for the
90$^{\circ}$ bond}. Upper signs: $t^{jq}$, lower signs: $t^{iq'}$.}
\label{tab2}
\begin{tabular}{c|ccc}
  & $p_{x}$ & $p_{y}$ & $p_{z}$   \\  \hline
0 & 0 &$\pm t_{0}$       & 0                \\
1 & 0 & $\pm t_{1}$ &  0               \\
$z$ &  $\pm t_{2}$ &0 & 0 \\
$x$ & 0 & 0 & $\pm t_{2}$ \\
$y$ & 0 & 0 & 0 \\ \hline
\end{tabular}
\end{table}
\begin{table}
\caption{The Cu--O--O hopping
matrix elements $T_{\alpha n}^{iq}$
{\it for the
90$^{\circ}$ bond}.
Upper signs: $T^{iq}$, lower signs: $T^{jq'}$.}
\label{tabn1}
\begin{tabular}{c|ccc}
  & $p_{x}$ & $p_{y}$ & $p_{z}$   \\  \hline
0 & $\pm t_{0}t_3/\epsilon_{p_x}$
& $\mp t_0 t_4/\epsilon_{p_x}  $     & 0         \\
1 & $\mp t_{1}t_3/\epsilon_{p_x}$
&     $\pm t_{1}t_4/\epsilon_{p_x}$   & 0          \\
$z$ & $\pm t_2t_4/\epsilon_{p_y}$ & $\mp t_{2}t_3/\epsilon_{p_y}$ &0 \\
$x$ & 0 & 0 & 0 \\
$y$ & 0 & 0 & $\pm t_{2}t_{5}/\epsilon_{p_z}$ \\ \hline
\end{tabular}
\end{table}
\begin{table}
\caption{The Cu--O--O hopping matrix elements $T_{\alpha n}^{iq}$ {\it for the
90$^{\circ}$ bond}. Upper signs: $T^{jq}$, lower signs:
$T^{iq'}$.}
\label{tabn2}
\begin{tabular}{c|ccc}
  & $p_{x}$ & $p_{y}$ & $p_{z}$   \\  \hline
0 & $\pm t_0 t_4/\epsilon_{p_y} $
& $\mp t_0t_3/\epsilon_{p_y}$ &0                \\
1 & $\pm t_1t_4/\epsilon_{p_y}$ & $\mp t_3t_{1}/\epsilon_{p_y}$ &  0
  \\
$z$ &  $\mp t_2t_3/\epsilon_{p_x}$ &
$\pm t_2t_4/\epsilon_{p_x}$ & 0 \\
$x$ & 0 & 0 & $\pm t_{2}t_{5}/\epsilon_{p_z}$ \\
$y$ & 0 & 0 & 0 \\ \hline
\end{tabular}
\end{table}

\end{document}